\documentstyle[pra,preprint,aps]{revtex}    

\tighten

\begin{document}

\draft

\title{NONLOCAL AND LOCAL GHOST FIELDS IN QUANTUM CORRELATIONS}
\author{Krzysztof W\'odkiewicz}
\address{Center for Advanced Studies, Department of Physics and
Astronomy, \\
University of New Mexico, Albuquerque, New Mexico 87131,USA \\
and \\
Instytut Fizyki Teoretycznej, Uniwersytet Warszawski
Warszawa 00-681, Ho\.za 69, Poland \cite{aaa}}
\date{\today}
\maketitle
\begin{abstract}
Einstein Podolsky Rosen quantum correlations are
discussed from the perspective of a ghost field introduced by Einstein.
The concepts of ghost field, hidden variables, local reality and the
Bell inequality are reviewed. In the framework of the correlated
singlet state of two spin-$\frac{1}{2}$ particles, it is shown that
quantum mechanics can be cast  in  a way that has the form of either  a
nonpositive and local ghost field or a positive and nonlocal ghost
field.

\end{abstract}
\pacs{PACS numbers:  03.65.Bz}

\section{Introduction}
     Despite the spectacular success of quantum mechanics as a
     fundamental theory and as a superb calculational tool for all
modern physics, the scientific literature continuously carries  a
substantial number of papers devoted to the foundations and the
interpretation of the theory.  The introduction of probability in the
sense used in quantum mechanics has been the key issue of debates.
Since its early days the probabilistic interpretation of quantum
phenomena has had its opponents, and Einstein in a letter to Born
expressed the view \cite{pais1}:

\vskip0.3cm {\sl The theory produces a good deal but hardly brings us
closer to the secret of the OLD ONE. I am at all events convinced that
{\sl He} does not play dice.} \vskip0.3cm

This statement is the best known example of the open criticism of the
probabilistic interpretation of quantum mechanics.

In his fundamental paper devoted to the quantum theory of spontaneous
and stimulated processes by radiation, Einstein expressed his
well-known discomfort about the probabilistic description of
spontaneous emission in the following words \cite{einstein}:
\vskip0.3cm {\sl The weakness of the theory lies on the one hand in the
fact that it does not get us any closer to making the connection with
wave theory; on the other hand, that it leaves the duration and
direction of the elementary process to 'chance'.} \vskip0.3cm

   In making the connection between waves and particles, Einstein noted
   that the emission of radiation (outgoing radiation) is directional.
In the same paper he wrote:

\vskip0.3cm {\sl Outgoing radiation in the form of spherical waves does
not exist.} \vskip0.3cm

     In the early 1920's, Einstein, in his unpublished speculations,
proposed the idea of a  ``Gespensterfeld'' or a ghost field
 which determines the probability for a light-quantum to take a
definite path.  In these speculations, the ghost field gives the
relation between a wave field and a light-quantum by triggering the
elementary process of spontaneous emission.  The directionality of the
elementary process is  fully described by the will (dynamical
properties) of the ghost field (see Fig. \ref{fig1}).

     The probabilistic interpretation of quantum mechanics as developed
     by Born has been considerably influenced by the idea of the ghost
field \cite{born}.  Born realized that unlike the electromagnetic
field, the Schr\"odinger wave function $\Psi$ has no direct physical
reality.  Only if one interprets $|\Psi|^2$ or the ghost field as a
real probability, do all the weaknesses of the  interpretation
disappear because:
\begin{equation} {\cal GHOST \ FIELD}
\longrightarrow |\Psi|^2
\end{equation}

In a  paper presented at Oxford,  Born  expressed the opinion that
\cite{born2}:
\vskip0.3cm

{\sl But, of course, anybody dissatisfied with these ideas
may feel free to assume that there are additional parameters not yet
introduced  into the theory which determine the individual event.}
\vskip0.3cm

\noindent These additional parameters called, accordingly, hidden
variables and denoted by $\lambda$, have been introduced in order to
remove from quantum mechanics the indeterminacy introduced by the Born
interpretation of the wave function.  According to hidden variables
theories, all physical  observables such as, for example the position
and the momentum of a particle, the spin or the electric field  are
entirely deterministic, i.e, are objective realities given by functions
$O(\lambda)$. Due to the assumed unobservable nature of the hidden
parameters $\lambda$, only the following average

\begin{equation}
\langle O\rangle _{HV} = \int d \lambda \ P(\lambda)\ O(\lambda)
\end{equation}
can be measured in realistic experimental situations.  The distribution of the
as yet undiscovered hidden variables is positive (a well-behaved probability
distribution) and normalized

\begin{equation}
P(\lambda) \ge 0 {\rm \ and\ } \int d\lambda\ P(\lambda) = 1 .
\end{equation}
In this interpretation the hidden parameters would be equivalent to a ghost
field of some yet undisclosed nature, i.e.,
\begin{equation}
{\cal GHOST \ FIELD} \longrightarrow  P(\lambda)
\end{equation}

In 1935, Einstein, Podolsky, and Rosen (EPR) published their famous
paper \cite{epr} in which they argued that:
\vskip0.3cm {\sl
The wave function $\Psi$  does not provide a complete description of physical
reality.}
\vskip0.3cm

The need for a revision of the interpretation of the wave function was
strongly expressed in this paper. In fact, EPR questioned the
completeness of quantum mechanics, analyzing a correlated system of two
particles.  The EPR arguments can be best
summarized in  Bohm's version \cite{bohm1,bohm2} which uses
two correlated spin-$\frac{1}{2}$ particles as the EPR physical system.

Let us consider an initial physical system consisting of two distinguishable
spin-$\frac{1}{2}$ particles $a$ and $b$ in a spin-$0$ quantum state (a spin
singlet state). This system  dissociates to a pair of spatially separated
spin-$\frac{1}{2}$ particles. Let assume that the dissociation process
conserves the total spin of the system and that the separated particles are now
moving to opposite Stern-Gerlach detectors (Fig. \ref{fig2}). In the framework
of quantum mechanics the Bohm singlet wave function corresponding to the EPR
dissociating system  of two spin-$\frac {1}{2}$ particles $a$ and  $b$ is:
\begin{equation}\label{singlet}
|\Psi\rangle = { 1 \over \sqrt{2}}
\left(| + \rangle_a \otimes |-\rangle_b - | - \rangle_a \otimes
|+\rangle_b \right)
\end{equation}
where  $|+\rangle$ and  $|-\rangle$
are the two eigenstates of the Pauli matrix ${\hat \sigma}_z$ (quantum
operators are denoted with carets) corresponding to the two eigenvalues
$+1$ and $-1$.  After the dissociation the particles $a$ and $b$ move to
opposite directions without ever being disturbed in any way and various spin
components of each of these particles may by measured independently.

 The particle  $a$ is detected by a Stern-Gerlach apparatus which
can be oriented along  directions $\vec a$,  $\vec b$, and  $\vec c$
respectively.  The particle  $b$ is detected by a Stern-Gerlach apparatus which
can be oriented along the same three directions $\vec a$,  $\vec b$, and  $\vec
c$ respectively (Fig. \ref{fig2}).

 The EPR argument is based on the following observation
involving such measurements.  The Stern-Gerlach apparatus may record
that the spin $a$ is parallel or antiparallel to  $\vec a$, which one
can take to be the z-axis.  Let us assume that it has been detected as
parallel to  $\vec a$ meaning that we know that the value of the spin
$a$ is $ +1$. Knowing that the spin of particle $a$
is $+1$, common sense  based on an objective reality associated with
the spin (a singlet state consists of two antiparallel spins) predicts that the
second particle  $b$  has its spin $-1$. This means that  that the result of a
measurement carried on the second particle by a Stern-Gerlach apparatus with
the same orientation $\vec a$ will predict an antiparallel spin.

The same argument based on objective realities for spin variables can be
repeated for the orientations $\vec b$ and
$\vec c$ of the  Stern-Gerlach apparatus leading to the conclusion that
the values of the spin are known in the three directions which is in
violation of quantum mechanical properties of Pauli spin operators
projected along $\vec a$,  $\vec b$, and  $\vec c$ respectively.

Accordingly  EPR concluded that one should search for a deeper-lying theory in
which
the following two very important criteria are satisfied:

\begin{itemize}

 \item  {\bf The reality criterion}:  {\sl If without in any way
 disturbing the system, we can predict with certainty the value of a
physical quantity, then there exists an element of physical reality
corresponding to this physical quantity.}

 \item  {\bf The locality criterion}:   {\sl If at the time of
measurement... two systems no longer interact, no real change can take
place in the second system in consequence of anything that may be done
to the first system.}

\end{itemize}

The EPR  analysis of quantum correlations based on the reality and the
locality criteria have led to apparently contradictory conclusions with
quantum mechanics which do not allow for commuting objective realities
associated with the spin observables. EPR summarized their views in the
following statement:  \vskip0.3cm {\sl This simultaneous
}[predictability] {\sl makes the reality of} [the spins in Bohm's
version] {\sl dependent upon the process of measurement carried out on
the first system which does not disturb the second system in any way.
No reasonable definition of reality could be expected to permit this.}
\vskip0.3cm

If  the spin singlet  wave function (\ref{singlet})  of the two
particles is  measured by two Stern-Gerlach analyzers, in the
directions given by unit vectors  $\vec a$ and $\vec b$, the single
detection of the spins is denoted by $E(\vec a)$ and $E(\vec b)$, and
the joint simultaneous detection of the particles $a$ and $b$ is denoted by
$ E(\vec a; \vec b)$. These expectation values  for this wave function
(\ref{singlet}) are given by the following formulas:
\begin{eqnarray}\label{qepr}
E(\vec a)& =& \langle \Psi |{\hat \sigma}(\vec a) |\Psi \rangle \ = 0,
\nonumber \\ E(\vec b)& =& \langle \Psi |{\hat \sigma}(\vec b) |\Psi
\rangle \ =0, \nonumber \\ E(\vec a; \vec b)& =& \langle \Psi| {\hat
\sigma}(\vec a) \otimes {\hat \sigma} (\vec b) |\Psi \rangle = -{\vec
a}  \cdot {\vec b},
\end{eqnarray}
where ${\hat  \sigma(\vec a)}={\hat{\vec \sigma_a}}\cdot{\vec a} $ and
${\hat {\vec \sigma} }(\vec b)={\hat \sigma_b}\cdot{\vec b} $ are the
projections of the three spin-$\frac {1}{2}$  Pauli matrices  of the particle
$a$ and $b$ on the Stern-Gerlach apparatus orientations characterized by the
unit vectors  $\vec a$ and $\vec b$.

Quantum mechanics predicts that for parallel detectors we have a
perfect correlation:
\begin{equation} ({\hat \sigma}_{a  z}\otimes
{\hat \sigma}_{b  z})|\Psi\rangle = (\pm 1) (\mp1)  |\Psi\rangle = -
|\Psi\rangle
\end{equation}
i.e.,  the eigenvalue of ${\hat \sigma}_{a
z}$ is $\pm 1$ and as a result we have with probability one that the eigenvalue
of ${\hat \sigma}_{b z}$ is $\mp 1$.
 This perfect correlation between these two measurements permits one to
predict the  outcome of the second measurement with probability {\bf
one} without ever disturbing the second system.

On other hand in the framework of hidden variables or an unknown ghost
field these expectation values are  given by the following expression
\cite{redhead}:

\begin{eqnarray}\label{lhv} E(\vec a)& =&\int d\lambda_a \int
d\lambda_b \ P(\lambda_a;\lambda_b) \ \sigma (\vec a,\lambda_a) ,
\nonumber \\ E(\vec b)& =&\int d\lambda_a \int d\lambda_b
\ P(\lambda_a;\lambda_b) \ \sigma (\vec b,\lambda_b) , \nonumber \\
E(\vec a;\vec b)& =& \int d\lambda_a \int d\lambda_b
\ P(\lambda_a;\lambda_b) \ \sigma (\vec a,\lambda_a) \ \sigma (\vec
b,\lambda_b).  \end{eqnarray} In this formula  $P(\lambda_a;\lambda_b)$
describes a joint probability distribution of the ghost field
characterized by the hidden variables $\lambda_a$ and $\lambda_b$. The
objective realities of the spin variables are given by the
deterministic functions $\sigma (\vec a,\lambda_a)$ and $\sigma (\vec
b,\lambda_b)$.

The locality assumption leads to the property that  the distribution
function of the ghost field is independent of the surrounding
experimental setup and does not depend on what other observables are
measured simultaneously on the same system.

It is the purpose of this article to provide a quantum mechanical
description of the ghost field.   Using different descriptions and
formulations of quantum mechanics, we shall discuss the same EPR
correlation in different, though, equivalent forms of quantum theory.
In short, we shall allow quantum mechanics to speak for itself on such
issues as locality and the ghost field.  We believe that letting
quantum mechanics speak out via its different formulations  can provide
the most down-to-earth description of these fundamental and puzzling
issues.  As a result, our approach will be almost completely divorced from all
of its philosophical or speculative aspects.  Instead of this, we
shall focus our effort on the question:  How can we make quantum
mechanics look like a ghost field  and vice versa?  These
considerations demonstrate a much closer correspondence between the
probabilistic interpretation of quantum mechanics and hidden variables
or ghost fields. In our approach we shall show that  the wave function
$\Psi$ can always be replaced by a suitable quantum mechanical ghost
field $P_{qm}(\lambda_a;\lambda_b)$  with properties that defy the EPR
concepts of objective realities or locality.

\begin{equation} {\cal GHOST \ FIELD}
\longrightarrow |\Psi|^2 \longrightarrow P_{qm}(\lambda_a;\lambda_b).
\end{equation}

Expressing the formalism of quantum mechanics in a given and suitable
form  can provide a better understanding of the nature of the quantum
mechanical ghost field.  Instead of being vague in most of such
descriptions, the formalism of quantum mechanics can provide a
reasonable (if not ideal) description of quantum locality, reality,
nonlocality and their interrelations.

\section{Local Hidden variables and the Bell inequality}

  The concept of local realism as introduced by EPR is based on the
  fundamental assumption  that physical systems can be described by
  local objective properties that are independent of observation.
Starting with the EPR article,  questions devoted to local realism, to
measurements and the nonlocal character of quantum correlations, and to
hidden variables   have been raised, debated and analyzed both
theoretically and experimentally.  This debate has produced various
alternative theories to quantum mechanics.  These theories have been
considered over the years, hidden variables  theories being prominent
among these.

 From the theoretical point of view, a real breakthrough came after
 Bell discovered\cite{bell1,bell2} in 1965 that  certain classes of hidden
 variables theories limit the strength of spin or photon  correlations
 and that quantum mechanical correlations are not constrained by  these
 limits.  These fundamental limitations or constrains inhibited by
 local realism lead to the famous Bell inequality.

For the purpose of proving the Bell inequality we shall assume that
Stern-Gerlach apparatuses for the two particles are oriented along the
 three directions $\vec a$, $\vec b$ and $\vec c$ with angles $\vec a
 \cdot \vec b = \vec b \cdot \vec c= \vec c \cdot \vec a =
 \cos120^{\circ}$ ( see Fig. \ref{fig2}).

 In order to prove the Bell inequality we introduce (following Cover
 \cite{cover}) the following quantity built from objective spin
realities:

\begin{equation}\label{sigma}
 \Sigma(\lambda_a , \lambda_b)  = -\left( \sigma(\vec a,\lambda_a)
 +\sigma(\vec b,\lambda_a) +\sigma( \vec c,\lambda_a)\right) \left(
\sigma(\vec a,\lambda_b) +\sigma(\vec b,\lambda_b) +\sigma(\vec
c,\lambda_b)\right)
 \end{equation}
We shall restrict the range of these
objective realities to $ |\sigma(\vec i,\lambda_a) | \leq 1$ and  $
|\sigma(\vec i,\lambda_b) | \leq 1$ for all possible orientations
${\vec i} ={\vec a},{\vec b},{\vec c}$ of the Stern-Gerlach apparatus.
Because the spins are antiparallel we have with certainty that:

\begin{equation}\label{perfect}
\sigma(\vec a,\lambda_a) \cdot
\sigma(\vec a,\lambda_b)= \sigma(\vec b ,\lambda_a) \cdot \sigma(\vec
b,\lambda_b)= \sigma(\vec c,\lambda_a) \cdot \sigma(\vec c,\lambda_b)=
-1
\end{equation}
for all the three orientations  and for all the
possible hidden variables $\lambda_a$ and $\lambda_b$ forming the ghost
field that governs the statistical properties of the  local objective
realities representing the spin. This follows from the fact that if the
first Stern-Gerlach apparatus oriented at $\vec a$, $\vec b$ and $\vec
c$ detects the spin direction to be $\pm 1$, the second Stern-Gerlach
apparatus oriented as in Fig. \ref{fig2} has to detect the other spin
of the singlet state (\ref{singlet}) as  $\mp 1$. From the fact that
the objective realities are bounded by $\pm 1$ we note that
(\ref{sigma}) reaches a minimum value:
\begin{equation}
\Sigma(\lambda_a , \lambda_b) \geq \Sigma_{min}(\lambda_a , \lambda_b)=
-(1-1-1)(-1+1+1)=1
\end{equation}
{}From the fact that the spins are antiparallel i.e., (\ref{perfect}) holds, we
have the
following inequality involving the objective realities:
\begin{eqnarray}
3 - \sigma(\vec a,\lambda_a)\sigma(\vec b,\lambda_b)
-\sigma(\vec a,\lambda_a)\sigma(\vec c,\lambda_b) -\sigma(\vec
b,\lambda_a)\sigma(\vec c,\lambda_b)\nonumber \\ - \sigma(\vec
b,\lambda_a)\sigma(\vec a,\lambda_b) -\sigma(\vec
c,\lambda_a)\sigma(\vec a,\lambda_b) -\sigma(\vec
c,\lambda_a)\sigma(\vec b,\lambda_b)  \geq 1
\end{eqnarray}
multiplying this inequality by  $P (\lambda_a; \lambda_b)$, and integrating
over
the ghost field variables  we obtain the following Bell inequality:
\begin{equation}\label{bell}
 E(\vec a;\vec b) + E(\vec a;\vec c) +
E(\vec b;\vec c) \leq 1 ,
\end{equation}
where we have used the fact that correlations are symmetric i.e., for example
$E(\vec a;\vec b) = E(\vec b;\vec a)$.

In order to disprove a  theory based on local objective realities a
{\it series} of three  of different  correlation experiments
is required  (in our case correlations $\vec a$ $\vec b$,
$\vec a$ $\vec c$ and $\vec b$ $\vec c$ ).

What we have seen in the process of deriving the Bell inequality
(\ref{bell}) is that  one could provide a much more accurate
 description of locality, local reality and correlations  in terms of
(\ref{lhv}).  What is even more important in this inequality is the
fact that all these ideas could be put to experimental verifications.

For the three  orientations from Fig. \ref{fig2}  the  inequality
(\ref{bell}) reduces to
\begin{equation}\label{violation} 2
E(120^{\circ}) + E(240^{\circ}) \leq 1 .
\end{equation}
and for the
quantum correlation (\ref{qepr}) we have:
\begin{equation} -2
\cos120^{\circ}- \cos240^{\circ}= \frac {3} {2} \leq 1 .
\end{equation}
which is false and clearly violates the Bell inequality
(\ref{bell}) for the spin-$\frac {1} {2}$ singlet state given by
(\ref{singlet}).

These inequalities have helped to put a large class of hidden variables
theories to experimental tests \cite{clauser}.  Theories based on local realism
have all failed these tests. So far, all the experiments involving correlations
constrained by the Bell's inequality have been positive for quantum mechanics
and negative for various versions of local-realism \cite{rev94}.

If only one correlation experiment is performed the odds against
objective realities  are not very good.  Even if disproved in a series
of three experiments,  objective realities  leave a puzzling question.
If they come so close to the quantum predictions, perhaps there are
some elements of truth in their descriptions given by (\ref{lhv}).  If
so, which particular assumptions of  a local hidden variable theory are
in agreement or disagreement with quantum mechanics.  In order to
answer such a question, we shall describe in the following sections,
genuine quantum correlations in a form which bears a lot of
similarities to the objective reality description.  Once quantum
mechanics has   been formulated in the form expressed by
Eq.~(\ref{lhv}) a genuine effort can be made to address such issues  as
hidden variable  objective realities and locality.  Once this relation
is established, the question why quantum mechanics violates the Bell
inequality can be posed and investigated.

\section{Local Quantum Ghost Field}

The classical Malus' Law predicts an attenuation of an incident
polarized light beam with intensity $I_0$, passing through a linear
polarizer to obey $I=I_0 \cos^2\alpha$.  This attenuation depends on the
relative angle $\alpha$ between the polarization direction $\vec n$ of
the incoming wave and  the orientation $\vec a$ of the polarizer, i.e,
$\cos \alpha = \vec n \cdot \vec a $.  If the incoming light beam
consists of a statistical mixture of polarized light, the incoming
intensity  is:

\begin{equation}\label{malus}
\langle I \rangle = \int d \vec n
\ P_{\it cl}(\vec n) \cos^2\alpha .
\end{equation}%
In this formula, the integration is over all possible angles  of the random
polarization direction $\vec n$  and  the classical distribution function
$P_{\it cl}(\vec n)$ characterizes the statistical properties of the   incident
light beam polarization.

In quantum mechanics a similar Malus' Law holds for spin-$\frac{1}{2}$
particles detected by a Stern-Gerlach apparatus. Let  assume that  the
spin of the  particle $a$ is oriented along an arbitrary unit
direction  ${\vec n}_a$.  The quantum expectation value of such a
spin, if  detected by the Stern-Gerlach apparatus oriented in the
direction $\vec a$ is:
\begin{equation}
 \langle {\vec n}_a |{\hat \sigma}(\vec a) |{\vec n}_a \rangle \ =
 {\vec n}_a \cdot {\vec a} =\cos \alpha ({\vec a},{\vec n}_a)
\end{equation}
In this expression  $\alpha ({\vec a},{\vec n}_a)$ is the relative
 angle between the  orientation of the detected spin state $ |{\vec
n}_a \rangle$ and the Stern-Gerlach apparatus ${\vec a}$. The
corresponding expectation value of the spin-$\frac{1}{2}$ projection
operator $ {\hat P}(\vec a) = \frac {1}{2} (1 +{\hat \sigma}(\vec a))
$ is:
\begin{equation} \label{malus2}
\langle {\vec n}_a |{\hat P}(\vec a) |{\vec
n}_a \rangle = \frac {1}{2} (1 +{\vec n}_a \cdot {\vec a} ) = \cos^2
\frac {\alpha}{2}
\end{equation}
is just the quantum version of the
Malus' Law for spin. Light (\ref{malus}) and spin-$\frac{1}{2}$ (\ref{malus2})
differ in this formulation by a factor $\frac{1}{2}$ in the relative angle
involved in the Malus law.

Following the classical Malus' Law for an  unpolarized light beam
(\ref{malus}), one can write the following quantum spin correlation
function  for an arbitrary  quantum state  of the two spin-$\frac{1}{2}$
particles detected by the Stern-Gerlach apparatus:
\begin{equation}\label{qmalus}
 E(\vec a; \vec b) = \int d {\vec n}_a\ \int d {\vec n}_b\ P({\vec
n}_a; {\vec n}_b) \ \cos \alpha({\vec a},{\vec n}_a) \ \cos
\alpha({\vec b},{\vec n}_b) .
\end{equation}
where the distribution function $ P({\vec n}_a; {\vec n}_b) $ describes the
quantum mechanical ghost field  for an arbitrary state $|\Psi \rangle$ of the
two spin-$\frac{1}{2}$ particles.  This quantum mechanical expectation
value (\ref{qmalus}) has a remarkable similarity to the hidden variable
equation (\ref{lhv}) if the following correspondence is made
\cite{kw1,kwmos1}:
\begin{equation} \begin{array}{ccccccc}
{\vec n}_a
& \leftrightarrow & \lambda_a & \ \ {\rm and } \ \ & {\vec n}_b
&\leftrightarrow & \lambda_b  \\ \sigma(\vec a, \lambda_a) &
\leftrightarrow& \cos\alpha({\vec a},{\vec n}_a) & \ \ {\rm and } \ \
& \sigma(\vec b, \lambda_b) & \leftrightarrow &\cos\alpha({\vec
b},{\vec n}_b).
\end{array}
\end{equation}
i.e., the  $\cos\alpha({\vec a},{\vec n}_a)$ and $\cos \alpha({\vec b},{\vec
n}_b ) $ are now the directional cosine functions between the directions of the
polarizers $\vec a$ and $\vec b$ and the ``hidden-variable'' directions
$\lambda_a={\vec n}_a $ and $\lambda_b ={\vec n}_b $  averaged with
respect to a local, i.e., Stern-Gerlach apparatus independent  ghost
field $P(\lambda_a ; \lambda_b) = P({\vec n}_a; {\vec n}_b) $.

 It is then clear that the quantum mechanical formula (\ref{qmalus})
has the form of a hidden variable theory with the local spin realities
given by $\cos(\alpha({\vec a},{\vec n}_a)$ and $\cos \alpha({\vec
b},{\vec n}_b) $, and with a local ghost field represented by the
distribution $ P({\vec n}_a;{\vec n}_b)$.

For the EPR  correlated system of the two spin-$\frac{1}{2}$ systems,
it is  natural just to take the probability distribution in the form:
\begin{eqnarray}\label{lgh}
P({\vec n}_a;{\vec n}_b) =
\frac{1}{4\pi}\ \delta^{(2)}({\vec n}_a +{\vec n}_b) \nonumber \\ \int
d {\vec n}_a \int d {\vec n}_b \ P({\vec n}_a;{\vec n}_b) =1.
\end{eqnarray}
where the two-dimensional Dirac delta function
indicates that the direction ${\vec n}_a$ is antiparallel to the
direction ${\vec n}_b$ ie., ${\vec n}_a =-{\vec n}_b$.

 The marginal properties of this distribution represent just an
isotropic  uniform distribution of the spin directions on the sphere:
\begin{equation}\label{marginals1}
\int d {\vec n}_b \ P({\vec
n}_a;{\vec n}_b) = \frac {1} {4 \pi} \ \ \ \ {\rm and} \ \ \ \int d
{\vec n}_a \ P({\vec n}_a;{\vec n}_b) = \frac {1} {4 \pi}.
\end{equation}

{}From these relations we conclude that individual ghost fields of the
spins correspond to a uniform distribution of orientations. This leads to
single spin averages   equal to zero, i.e.,:
\begin{eqnarray}
 E(\vec a) =  \frac {1} {4 \pi} \int d {\vec n}_a \ {\vec n}_a \cdot
 \vec a =0, \nonumber \\ E(\vec b) = \frac {1} {4 \pi} \int d {\vec
 n}_b \  {\vec n}_b \cdot \vec b =0.
\end{eqnarray}

On the other hand a direct integration of the Malus correlation
(\ref{qmalus}) with the antiparallel distribution function given by Eq.
(\ref{lgh}) leads to:
\begin{eqnarray}
E(\vec a, \vec b) =
\frac{1}{4\pi} \int d {\vec n}_a\int d {\vec n}_b \ \delta^{(2)}({\vec
n}_a +{\vec n}_b) \
({\vec n}_a \cdot \vec a )\ ({\vec n}_b \cdot \vec b ) \nonumber \\
 = -  \frac{1}{4\pi} \int d {\vec n}_a \ (\vec a \cdot {\vec n}_a )\
(\vec b \cdot{\vec n}_a )= - \frac {1}{3}{\vec a}\cdot{\vec b}.
\end{eqnarray}
This result  does not violate the Bell inequality
(\ref{bell}) and is off from the quantum mechanical result (\ref{qepr})
by a factor of $3$. This can be simply fixed by a change of the
$\frac{1}{4\pi}$-prefactor in (\ref{lgh}) into $\frac{3}{4\pi}$. But
this change has tremendous consequences for the normalization of the
distribution function (\ref{lgh}). In order to preserve the
normalization we write:
\begin{equation}\label{qlgf}
P_{qm}({\vec
n}_a;{\vec n}_b) = \frac{3}{4\pi}\delta^{(2)}({\vec n}_a +{\vec n}_b) -
 \frac {2} { (4\pi)^2}.
\end{equation}
We check that indeed with this
change:
\begin{equation}
 \int d {\vec n}_a \int d{\vec n}_b
\left(\frac{3}{4\pi}\delta^{(2)}({\vec n}_a +{\vec n}_b) -
 \frac {2} { (4\pi)^2}\right)= \frac {3} {4 \pi} 4\pi - \frac {2} {
 (4\pi)^2} (4\pi)^2 = 1.
\end{equation}
We calculate the marginals:
\begin{eqnarray}\label{marginals}
\int d {\vec n}_a
\left(\frac{3}{4\pi}\delta^{(2)}({\vec n}_a +{\vec n}_b) -
 \frac {2} { (4\pi)^2} \right) = \frac {3} {4 \pi}- \frac {2} {
(4\pi)^2} 4\pi = \frac {1} {4 \pi}, \nonumber \\ \int d {\vec n}_b
(\frac{3}{4\pi}\delta^{(2)}({\vec n}_a +{\vec n}_b) -
 \frac {2} { (4\pi)^2}) = \frac {3} {4 \pi}- \frac {2} { (4\pi)^2} 4\pi
= \frac {1} {4 \pi}.
\end{eqnarray}
Clearly the function (\ref{qlgf}) has the correct marginals (\ref{marginals}),
reproduces the right quantum mechanical prefactor and yet is not positive!
The appearance of such a nonpositive probability distribution function (called
for this reason a quasi-distribution function) is a typical quantum property,
whenever a probabilistic phase-space is constructed for noncommuting quantum
observables.
The first nonpositive quasi-probability distribution was introduced into
quantum mechanics by Wigner in 1932 (\cite{wigner}) for the position and the
momentum phase-space.   The  quasi-distribution function (\ref{qlgf})
corresponds to a Wigner-type distribution function for the spin-$\frac{1}{2}$
phase-space variables. In fact it  is possible to study and  derive rigorously
\cite{kw2,kwmos2} all kinds of quantum mechanical spin-$\frac{1}{2}$
quasi-distribution functions with (\ref{qlgf}) being just an example
corresponding to the quantum ghost field of the EPR  singlet state
(\ref{singlet}).

The quantum Malus Law (\ref{qmalus}), if applied to the joint
correlations involving two Stern-Gerlach detectors, has the formal
structure of a hidden variable theory, with the hidden parameters
represented by "hidden directions " ${\vec n}_a$ and $ {\vec n}_b$.
There the analogy ends because the quantum distribution corresponding
to the local  ghost field is a non-positive function that leads to the
failure of the Bell's inequalities for such a correlated state (see
Fig. \ref{fig3}). Using the Malus form of spin correlations we have:

\begin{equation}
{\cal LOCAL \ \ GHOST \ FIELD} \Longrightarrow
 \frac{3}{4\pi}\delta^{(2)}({\vec n}_a +{\vec n}_b) - \frac {2} {
 (4\pi)^2}.
\end{equation}

\section{Nonlocal quantum ghost field}

     In this section we are going to show that EPR correlations can be
described by a quantum nonlocal ghost field  similar to the one used in
a hidden-variables theory.

We can rewrite the spin correlation function (\ref{qepr}) in terms of a
ghost field if we use the following two integral identities for the
Pauli matrices matrices:

 \begin{equation}
{\hat \sigma}(\vec a) = \int d \lambda_a \lambda_a
\delta (\lambda_a - {\hat \sigma}(\vec a)), \ \ {\hat \sigma}(\vec b) =
\int d \lambda_b \lambda_b \delta (\lambda_b - {\hat \sigma}(\vec b)).
\end{equation}
Introducing these identities into (\ref{qepr})we
obtain:

\begin{equation}
E(\vec a; \vec b) = \langle \Psi| {\hat \sigma}(\vec
a) \otimes {\hat \sigma} (\vec b) |\Psi \rangle = \int d\lambda_a \int
d\lambda_b \ P({\vec a},\lambda_a;{\vec b}, \lambda_b) \ \lambda_a
\ \lambda_b .
\end{equation}
where the distribution function is given by the following quantum mechanical
average:
\begin{equation}\label{ccd}
P({\vec a},\lambda_a ;{\vec b}, \lambda_b)
= \langle \Psi| \delta (\lambda_a - {\hat \sigma} (\vec a)) \otimes
 \delta (\lambda_b - {\hat \sigma} (\vec b))|\Psi \rangle.
\end{equation}
With the help of this distribution function we have rewritten the
quantum mechanical correlation function in a form which has remarkable
similarities to the  correlation function given by Eq.~(\ref{lhv}).
Because the spin operators ${\hat \sigma}(\vec a)$ and ${\hat
\sigma}(\vec b)$  have eigenvalues equal to $+1$ or $-1$, i.e., can
represent only ``parallel '' (``p'' ) or ``antiparallel'' (``a'')
outcomes,  $\lambda_a $ and $\lambda_b $ can take only values equal to
$+1$  and $-1$.  The bivalued distribution given by Eq.~(\ref{ccd}) is
positive  everywhere and normalized:  \begin{equation} \int d\lambda_a
\int d\lambda_b \ P({\vec a},\lambda_a;{\vec b}, \lambda_b)= \langle
\Psi | \Psi \rangle =1.  \end{equation} This function  depends on the
polarization directions $\vec a$ and $\vec b$.  The distribution
function which depends on the orientation  $\vec a$ of the first
Stern-Gerlach apparatus and on the orientation $\vec b$ of the second
(possibly even remote) Stern-Gerlach apparatus is nonlocal. Because of
this property we shall  call an analyzer-dependent distribution
function a nonlocal distribution function. The nonlocality of this
distribution function makes the Bell's inequality void, because in
order to obtain this inequality the existence of a universal, local
(polarization independent) distribution in the parameters $\lambda_a $
and $\lambda_b $  (hidden parameters in this case) like that appearing
in (\ref{lhv}) was essential \cite{mermin}. Quantum mechanics tells us that if
we insist on a distribution of the form given by Eq.~(\ref{ccd}), we can
do it but only under the condition that the statistical distribution of
the ghost field with the parameters  $\lambda_a $ and $\lambda_b $  is
nonlocal \cite{kwjoe}.

This joint distribution function (\ref{ccd}) can be written in the
following form (Bayes theorem):

\begin{equation}
P(\lambda_a; \lambda_b) = P(\lambda_a| \lambda_b)
P(\lambda_b)
\end{equation}
where the distribution $  P(\lambda_a|\lambda_b)$ is the conditional of the
event  $\lambda_a $ to occur under the condition that $\lambda_b $ has
occurred.     The distribution $ P(\lambda_b)$ is a one-fold marginal of
(\ref{ccd}) and
is:
\begin{equation}
P({\vec b}, \lambda_b) = \int d\lambda_a  \ P({\vec
a},\lambda_a;{\vec b}, \lambda_b) = \langle \Psi| \delta (\lambda_b -
{\hat \sigma} (\vec b))|\Psi \rangle = \frac{1}{2}.
\end{equation}
This means the the one-fold marginal is local (apparatus independent)
and that the outcomes are equally probable ie., $P(\lambda_b=+1)=\frac
{1} {2}$ and $P(\lambda_b=-1)=\frac {1} {2}$.

In order to elucidate the nonlocal distribution further we shall
rewrite the nonlocal conditional distribution function in the following
matrix form:
\begin{equation}
P(\lambda_a| \lambda_b) =\left(
\begin{array}{cc} P(+1|+1 ) &P(+1|-1) \\ P(-1|+1) & P(-1|-1)
\end{array} \right)
\end{equation}
{}From Eq.(\ref{ccd}) and the form  of the EPR singlet state we obtain
that the conditional probabilities for the particular transitions are
\begin{equation}\label{qnlgf}
P(\lambda_a| \lambda_b) =\left(
\begin{array}{cc} \sin^2 \frac {\alpha}{2}  &\cos^2 \frac {\alpha}{2}
 \\ \cos^2 \frac {\alpha}{2} & \sin^2 \frac {\alpha}{2} \end{array}
\right)
\end{equation}

     This result shows that one can regard the EPR correlations as just
correlations of two sequences of random numbers jumping between values
$+1$ and $-1$  (``p'' and ``a'' answers) for polarization measurements
performed with linear analyzers. The distribution $P(\lambda_a|
\lambda_b)$ is the conditional of the event $\lambda_a$ (``p'' or
``a'') to occur under the condition that $\lambda_b$ (``a'' or ``p'')
has occurred.

 This positive and nonlocal distribution leads to a simple statistical
interpretation of the spin transitions and of the violation of Bell's
inequality in terms of random numbers  $\pm 1$ for the variables
$\lambda_a$ and $\lambda_b$.  The quantum mechanical average in this
case is represented by an ensemble average of two sequences of random
numbers ``a'' and ``p''.  The random character of these variables can
be applied to the description of the EPR correlations measured by two
polarizers.  To each polarizer there corresponds a sequence of random
variables denoted by $\lambda_a$ and $\lambda_b$.  These are the only
possible outcome of the experiment.  On each single polarizer the
outcomes are completely random and the ``p'' and ``a'' answers occur
with equal probability ($\frac {1}{2}$ in this case).  The nonlocality
of the EPR correlations shows up in the fact that these two perfectly
random sequences (on the first and the second analyzer) are correlated
and the correlations are given by Eq.~(\ref{qepr}). These formulas
predict that the EPR wave function can be understood as a nonlocal
correlation between two random sequences $\lambda_a = (p, a, a, \ldots )$
and $\lambda_b = (a, a, p, \ldots )$.  The nonlocality of these
correlations follows from the fact that whenever $\lambda_b = p$ or
$\lambda_b = a$ on the polarizer $\vec b$, we must have $\lambda_a = p$
or $\lambda_a = a$ on the analyzer $\vec a$ with the probability $\sin^2
\frac {\alpha}{2}$, i.e., the outcomes on $\vec a$ (possibly even a
remote analyzer) are determined by the outcomes on the analyzer $\vec
b$.

For antiparallel correlations $\alpha =0$ and the conditional matrix
has the following form:
\begin{equation}
 P(\lambda_a| \lambda_b)
=\left( \begin{array}{cc} 0 &1 \\ 1 & 0 \end{array} \right)
\end{equation}
This ghost field correspond to the following two
outcomes at the Stern-Gerlach analyzers:
\begin{equation} \lambda_b = \left(
\begin{array}{c}
 p \\ a \\ a \\ \vdots \end{array} \right) \ \ \ \Longrightarrow \ \ \
\lambda_a =\left( \begin{array}{c}
 a \\ p \\ p \\ \vdots \end{array} \right).
\end{equation}
We see that the outcomes correspond to two random sequences with perfect
correlation corresponding to $ a \leftrightarrow p$ with conditional
probability one and $a \leftrightarrow a$ and $p \leftrightarrow p$
with conditional probability zero.

 Let us illustrate, using these random sequences, the violation of
 Bell's inequality.
 Let us assume that in the first series of experiments we set $\alpha =
120^{\circ}$ and $\alpha = 240^{\circ}$  .  According to the formulas
(\ref{qnlgf}) we have for these angles:
\begin{equation}
P(\lambda_a| \lambda_b) =
\left( \begin{array}{cc} \frac {3} {4}  & \frac {1} {4}  \\
\frac {1} {4}  & \frac {3} {4}
\end{array} \right).
\end{equation}
This means that  with a 75\% confidence we shall have that the outcomes
on $\vec a$ will be the same as the outcomes on $\vec b$ and  a
25\% confidence that the outcomes are different. This leads to the spin
correlation:
\begin{eqnarray}
E(120^{\circ}) = E(240^{\circ})=
\sum_{\lambda_a=\pm 1 \lambda_b=\pm 1} \lambda_a \ \lambda_b
\ P(\lambda_a |\lambda_b)\ \frac{1}{2} \nonumber \\ +1\cdot 0.75
-1\cdot 0.25 = - \cos (120^{\circ}) = \frac {1} {2}
\end{eqnarray}
We already have shown (\ref{violation}) that these correlation  do violate
the Bell inequality (\ref{bell}). The nonlocal ghost field  contradicts the
idea of objective realities because the probability of reproducing a sequence
on the polarizer $\vec a$ if the sequence on the polarizer $\vec b$ is known is
nonlocal if the ghost field distribution  depends on the relative orientation
$\alpha$ of the polarizers.  This is how the EPR quantum nonlocal
probabilities  violate local realism ( Fig. \ref{fig4}):
\begin{equation}
{\cal NONLOCAL\ GHOST \ FIELD } \Longrightarrow
P(\lambda_a |\lambda_b).
\end{equation}

In fact it is easy to give an example of a ghost field that will not
violate the Bell inequality in this case. If we select:

\begin{equation}
P(\lambda_a| \lambda_b) =\left( \begin{array}{cc}
\frac {5} {12}  & \frac {7} {12}  \\ \frac {7} {12}  & \frac {5} {12}
\end{array} \right)
\end{equation}
we obtain that:
\begin{equation}
E(120^{\circ}) = +1\cdot \frac {5} {12} -1\cdot \frac {7} {12} = -
\frac {1} {3} \cos 120^{\circ}
\end{equation}
which does not violate
the Bell inequality.  The quantum nonlocal ghost field predicts in this
case that the outcomes on the two Stern-Gerlach detectors is the same with a
75\%
confidence while hidden variables will lead to $\frac {5}{12}\approx
41.6 \%$ confidence. The quantum correlations induced by the nonlocal
quantum ghost (\ref{qnlgf} ) are "stronger" compared to the one obtained in the
frame work of local objective realities (\ref{lhv}).

\section{Conclusions}

 The two different  representations of quantum mechanics given by a
 local and nonpositive ghost field and by a nonlocal and positive ghost
 field are equivalent descriptions of spin-$\frac {1}{2}$ correlations
 and both  have a hidden-variable look-a-like form.  These descriptions
are different in form, but are really equivalent to the standard
formulation of quantum mechanics.  Instead of spin observables,
projection operators and quantum states, the quantum ghost field
description of the EPR correlations involves local (or nonlocal),
positive (or nonpositive) distributions for random spin orientations
satisfying (or not satisfying) the Malus law.
The relations (\ref{qlgf}) and (\ref{qnlgf}) show that the
violation of Bell's inequality  can be due either to a nonpositive and local or
to a positive and nonlocal distribution function.

     We have exhibited in this paper the two extreme cases which show
     that quantum mechanics is equivalent to a hidden-variable theory
with nonpositive probabilities or to a hidden-variable theory with
nonlocal distribution functions.  Which view we adopt is quite
irrelevant because the two pictures are equivalent and represent simply
different aspects of the same {\bf quantum mechanical reality}.

\section*{Acknowledgments} The author is indebted to Prof. P. L. Knight
for the encouragement to write in this form my thoughts about the ghost
field, which I  presented for the first time at Imperial College in
1986. Over these years I have benefited enormously discussing with I. Bia\l
ynicki-Birula, C. Caves, T. Cover, J. H. Eberly, G. Herling,  P. Meystre, G.
Milburn, P. Milonni, K. Rz\c a\.zewski, M. O. Scully, J. T\"oke and A.
Zeilinger.

\section*{vita} Krzysztof W\'odkiewicz is professor at the Institute of
Theoretical Physics of the University of Warsaw in Poland and a
research professor at the Center for Advanced Studies and the
Department of Physics of the University of New Mexico, USA. His
scientific interests include quantum optics, radiation theory, stochastic
processes, foundations of quantum mechanics.

\begin{figure}
\caption[f1]{\label{fig1}
According to the Born interpretation of the wave function $\Psi$, the time and
direction of the spontaneously emitted light-quantum (from an excited state to
the ground state) is left to `chance'.  In order to overcome the `chance' and
the probabilistic nature of such a spontaneous emission act, Einsten speculated
in the early 1920's that a ghost field of a yet unspecified nature determines
the probability for the spontaneously emitted light-quantum to take a definite
path.}
\end{figure}

\begin{figure}
\caption[f2]{\label{fig2}
A system of two distinguishable spin-$\frac{1}{2}$ particles $a$ and $b$ with
total spin = 0 , dissociates into  a pair of spatially separated particles.
Spin components of each of these particles are then measured independently by
Stern-Gerlach detectors. The EPR joint correlations are detected by two
Stern-Gerlach apparatuses oriented along the three directions ${\vec a}$,
${\vec b}$ and ${\vec c}$.}
\end{figure}

\begin{figure}
\caption[f3]{\label{fig3}
 Spin correlations in the EPR system of two spin-$\frac{1}{2}$ particles can be
described to a local objective ghost field that determines the probability of
the individual spin orientations. In quantum mechanics this ghost field is
local and not positive and is given as $P_{qm}({\vec n}_a;{\vec n}_b)=
\frac{3}{4\pi}\delta^{(2)}({\vec n}_a +{\vec n}_b) - \frac {2} {(4\pi)^2}$.
The negative regions of the ghost field, denoted by $-$, make the Bell
inequality void in this case.}
 \end{figure}

\begin{figure}
\caption[f4]{\label{fig4}
  Spin correlations in the EPR system of two spin-$\frac{1}{2}$ particles can
be described by  a nonlocal objective ghost field that determines the
probability of the individual spin orientations. In quantum mechanics this
positive and nonlocal, appartus dependent, ghost field is given as  $
P(\lambda_a |\lambda_b).$ The nonlocality of the ghost field  makes the Bell
inequality void in this case.}
 \end{figure}

\end{document}